# Catalytic activity of Al-Cu-Fe-Ni-Cr high entropy alloy

Yogesh Kumar Yadav[1], Mohammad Abu Shaz[1], and Thakur Prasad Yadav[1,2*]

[1]Department of Physics, Institute of Science, Banaras Hindu University Varanasi-221005, Uttar Pradesh, India

[2]Department of Physics, Faculty of Science, University of Allahabad, Prayagraj-211002, Uttar Pradesh, India

**Abstract:**

Magnesium hydride ($MgH_2$) is a promising material for hydrogen storage because of its abundance and beneficial properties, such as high storage capacity and cost-effectiveness under mild conditions. Despite of these benefits, $MgH_2$ unfavorable thermodynamics and kinetics make it difficult to use in real applications. In this work, the hydrogen storage properties of $MgH_2$ have been improved using Al-Cu-Fe-Ni-Cr high entropy alloy (HEA) based catalysts, which has been synthesized via mechanical alloying. The experimental findings show that the beginning desorption temperature of $MgH_2$ significantly lowered from 425°C to 180°C by adding 5 wt. % Al-Cu-Fe-Ni-Cr HEA in $MgH_2$. Moreover, the catalyst shows enhanced kinetics, attaining 7.3 wt. % hydrogen absorption in 3 minutes at 320°C with 15 atm hydrogen pressure, and ~5 wt. % desorption in 6 minutes at 320°C. These results highlight, how much lower its desorption temperature is than those of other well-known catalysts. Over a span of 25 cycles, $MgH_2$ catalyzed by Al-Cu-Fe-Ni-Cr HEA exhibits remarkable cyclic stability with negligible fluctuations (~ 0.05 wt. %). After a thorough characterization of the materials, a workable catalytic mechanism for HEA was proposed in light of the results.

*Corresponding author

# 1. Introduction:

The severity and frequency of heat waves, heavy precipitation, and droughts are consequences of rising global temperatures [1]. According to the IPCC, the last ten years have been the warmest on record, with the Earth temperature rising by 1.1°C to levels not seen in 125,000 years [2]. The main causes of this warming are human activity, specifically greenhouse gas emissions from burning fossil fuels [3]. According to WHO, 90% of the world population now breathes poor air, freshwater supplies are in danger, and glaciers are melting more quickly as a result [4-5]. A switch to renewable energy is necessary to overcome these issues. With its high energy density, carbon neutrality, and sustainability, hydrogen energy presents a viable way to address the world energy dilemma [6]. Because of its flammability, hydrogen energy presents storage and transportation issues despite its potential [7]. Solid-state storage, which uses materials with void designs for large hydrogen capacities, provides a safer and more affordable option to high-pressure or liquid approaches [8]. With a capacity of 7.6 wt. % (110 g $H_2$/L), $MgH_2$ is the industry leader [9]. Its slow kinetics and high reaction enthalpy (75 kJ $mol^{-1}$), which necessitate desorption temperatures over 350°C, restrict its use. Strategies including catalysis, nano confinement, and alloying have been investigated to address these problems; catalysts have shown tremendous promise for enhancing kinetics and thermodynamics because of their low cost and capacity to promote hydrogen molecule dissociation and recombination [10].

To maximize $MgH_2$ performance, researchers have experimented with several catalysts and alloying techniques in the quest to improve hydrogen storage methods. The catalytic effects of transition elements, such as Ti, V, Mn, Fe, and Ni, on $MgH_2$ were first studied by Liang et al., (1999) [11]. According to their research, Ti had the fastest absorption rates, followed in order by Mg–V, Mg–Fe, Mg–Mn, and Mg–Ni. In terms of desorption, MgH–V showed the quickest kinetics, followed at lower temperatures by MgH–Ti, MgH–Fe, MgH–Ni, and MgH–Mn. Interestingly, V and Ti were found to be catalysts that had better effects on the parameters of absorption and desorption, respectively. Zhou et al., (2013) conducted follow-up research that demonstrated the effectiveness of Ti-based intermetallic catalysts, including TiAl, in reducing desorption temperatures in comparison to elemental Ti [12]. They also mentioned that doping the catalyst with two or more metals could increase its surface activity. Cermak et al., (2011) offered more insights into catalytic effects by examining the effects of Ni, $Mg_2Ni$, and $Mg_2NiH_4$ on $MgH_2$ desorption characteristics [13]. According to their research, $Mg_2NiH_4$ had a much greater catalytic effectiveness than both pure Ni and non-hydrated intermetallic $Mg_2Ni$. Investigating the impacts of Ti-based man-sized additions on the hydrogen storage properties of $MgH_2$, Berezovets et al., (2022) found that the catalytic effect increased with nano-Ti < nano-$TiO_2$< $Ti_4Fe_2O_x$ [14]. Notably, the addition of $Ti_4Fe_2O_x$ resulted in a greater hydrogen storage capacity. The catalytic effectiveness of a new catalyst, the quasicrystal of $Al_{65}Cu_{20}Fe_{15}$, on the de/rehydrogenation characteristics of $MgH_2$ was reported by Pandey et al., (2017) [15]. The onset desorption temperature of $MgH_2$ was found to have significantly decreased, going from roughly 345°C (for $MgH_2$ that had been ball-milled) to roughly 215°C. They also reported a notable increase in the amount of hydrogen that could be stored, observing 6.0 wt. % at 250°C in just 30 seconds. Even at lower temperatures of 200°C and 150°C, where about 5.50 wt. % and 5.40 wt. % of $H_2$ were absorbed within 1 minute, respectively, enhanced rehydrogenation kinetics were also visible. Similarly, in 30 minutes, about 5 wt. % of $H_2$ was absorbed at 100°C. According to Pandey and colleagues, these findings showed some of the lowest rehydrogenation kinetics and desorption temperatures for $MgH_2$ that have been attained with any known catalyst.

$Cu_{0.15}Ni_{0.35}Co_{0.25}Fe_{0.25}$, a multicomponent transition metal alloy, was presented by Gupta et al., (2024) and functions as a catalyst to increase $MgH_2$ capacity to store hydrogen [16]. Their results showed that at 250°C and a plateau pressure of about 0.4 bar, about 6% of the hydrogen was desorbed, with a desorption enthalpy of about 76.74 kJ/mol $H_2$. Furthermore, quick absorption kinetics were noted, with 3.5 wt. % of hydrogen absorbed in 3 minutes at 250°C and a 5.9 wt. % saturation capacity reached in 10 minutes at 20 bar hydrogen pressure. This study emphasizes how multicomponent alloy catalysts might improve $MgH_2$ hydrogenation characteristics. Research has primarily focused on catalysts with three or fewer elements, known as medium entropy alloys. Nevertheless, research shows that the storage qualities of hydrogen are improved by the addition of additional components. Consequently, the application of multicomponent catalysts to enhance $MgH_2$ hydrogen sorption is growing. In the last 10 years, HEAs have drawn a lot of interest as possible hydrogen storage materials due to their distinct structures and characteristics [17–18]. Five or more elements are usually present in HEAs, with concentrations varying from 5 to 35 at. % [19-20].These alloys have outstanding catalytic activity and property adjustability, making them a possible substitute for conventional transition metal alloys [21]. They display distinctive characteristics like lattice distortion, slow diffusion, high entropy effects, and a cocktail effect [22-23]. Wang et al. (2023) used a variety of electrochemical and physical characterization techniques to examine the catalytic characteristics of HEAs and certify their performance [24]. A penta-element HEA (Pt-Pd-Au-Cu-Fe/C) that was synthesized and used as an anode catalyst in a genuine direct ethylene glycol fuel cell (DEGFC) stack was the subject of their investigation. The HEA showed noticeably higher ethylene glycol oxidation reaction (EGOR) activity than commercial Pt/C. The HEA catalyst oxidation current density was 0.65 A/mg, three times greater than that of commercial Pt/C (0.22

A/mg). The HEA also demonstrated outstanding long-term durability. With an open-circuit potential of 0.58 V and a peak power density of 17.63 mW/cm², the DEGFC with the HEA catalyst was twice as powerful as the Pt/C-based DEGFC (7.37 mW/cm²). These findings demonstrate how HEAs, which combine increased activity and durability, have the potential to be excellent catalysts for DEGFC applications. Results were reported by Edalati et al., (2020) on a Ti-Zr-Cr-Mn-Fe-Ni HEA whose composition included 95 wt. % C14 Laves phase [25]. The alloy they observed had quick kinetics and no need for activation treatment, proving its capacity to absorb and desorb 1.7% of hydrogen at ambient temperature. An equiatomic Ti-Zr-Nb-Cr-Fe HEA was studied by Floriano et al., (2021) for its hydrogen storage characteristics at different temperatures [26]. According to their research, this alloy has potential hydrogen storage properties because it can reversibly absorb and totally desorb 1.9 wt. % of hydrogen at 473 K. An investigation of the hydrogen storage behavior of a high-entropy Ti–Zr–V–Cr–Ni equiatomic intermetallic alloy was conducted by Kumar et al., 2022 [27]. They discovered that this alloy had a 1.52 wt. % hydrogen storage capacity, indicating its potential for use in hydrogen storage applications.

Additionally, studies have looked into the possibility of using HEAs as catalysts to improve $MgH_2$ capacity to store hydrogen. In a study by Wei et al., (2023), different concentrations of $(Ti-V-Zr-Nb)_{83}Cr_{17}$ HEA were added to $MgH_2$ as catalysts to create $(Ti-V-Zr-Nb)_{83}Cr_{17}$ (x = 3, 6, 10) composites [28]. According to their findings, the $MgH_2$-6 wt. % $(Ti-V-Zr-Nb)_{83}Cr_{17}$ composite shown exceptional hydrogen sorption capabilities at 573 K, quickly absorbing 5.3 wt. % $H_2$ in just 5 minutes. The composite maintained its effectiveness at 523 K, absorbing 4.0 wt. % $H_2$ in 60 minutes. Additionally, the composite showed quick desorption kinetics, releasing 6wt. % and 4.1 wt. % $H_2$ at 623 K and 573 K, respectively, in 60 minutes. Furthermore, the $MgH_2$-6 wt. % $(Ti-V-Zr-Nb)_{83}Cr_{17}$ composite demonstrated comparatively low activation energies for both desorption and absorption of hydrogen (70.6 kJ/mol $H_2$) and 90.5 kJ/mol $H_2$. These results highlight the enhanced catalytic capabilities of Ti-V-Nb-based HEAs on $MgH_2$ hydrogen storage characteristics. Wan et al., (2023) reported the fabrication of Fe-Co-Ni-Cr-Mn HEA loaded $MgH_2$ and examined its effect on $Mg/MgH_2$ capacity to store hydrogen [29]. They saw a significant drop in the dehydrogenation reaction's activation energy, which went from 151.9 - 90.2 kJ $mol^{-1}$. The $MgH_2$ – 5 wt. % HEA composite showed reliable hydrogen storage processes for at least 50 cycles, releasing 5.6 wt. % $H_2$ at 280 °C in 10 minutes and absorbing 5.5 wt. % $H_2$ at 150 °C in 0.5 minutes. The catalytic characteristics of HEA $Al_{20}Cr_{16}Mn_{16}Fe_{16}Co_{16}Ni_{16}$ and its leached variation on the de- and re-hydrogenation behaviors of $MgH_2$ were investigated by Verma et al., (2024) [30]. Under the influence of the leached HEA-based catalyst, they found that the onset desorption temperature of $MgH_2$ significantly decreased, from 360 °C to 338 °C. Furthermore, the composite showed good kinetics, desorbing ~5.4 wt. % in 40 minutes after absorbing ~6.1 wt. % of hydrogen in just 2 minutes at 300 °C under 10 atm hydrogen pressures. The storage capacity of $MgH_2$ catalyzed with the leached HEA decreased very slightly after 25 cycles of de/re-hydrogenation. Fe-Co-Ni-Cr-Mo HEA nanosheets catalytic influence on $MgH_2$ capacity for hydrogen storage was studied by Tao Zhong et al., (2023) [31]. They discovered that at 200°C, 9wt. % Fe-Co-Ni-Cr-Mo doped $MgH_2$ began to de-hydrogenate, and at 325°C, it discharged up to 5.89wt. % of its hydrogen in 60 minutes. With lower de/hydrogenation activation energy values than $MgH_2$, the composite could absorb 3.23 wt. % $H_2$ in 50 minutes at a low temperature of 100°C. Furthermore, after 20 cycles, the composite showed no discernible decrease in hydrogen capacity. To improve $MgH_2$ capacity for hydrogen storage, Chen Song et al., (2024) added Fe-Co-Ni-Cr-Ti HEA [32]. They found that the MgH$_2$-HEA composite initial desorption temperature had decreased, and that the dehydrogenation activation energy was significantly decreased by 51%. Promising catalytic effects were demonstrated by the composite, which took a much shorter time to absorb hydrogen than pure MgH$_2$.

Because of their distinct structural and compositional characteristics, HEAs have been shown in the aforementioned research to exhibit remarkable catalytic activity in augmenting MgH$_2$ hydrogen storage capacity. In particular, Cr-Fe-Ni-based HEAs have shown a notably better catalytic effect on MgH$_2$ hydrogen storage. Building on these discoveries, our research investigates the possibility of enhancing the processes of hydrogenation and dehydrogenation by adding Al and Cu elements to the alloy composition. This was accomplished by synthesizing the MgH$_2$ - 5 wt. % HEA and MgH$_2$ - 7 wt. % HEA system by designing and synthesizing a nanocrystalline Al-Cu-Fe-Ni-Cr (ACFNC) HEA, which was then added to MgH$_2$ using a high-energy ball mill. We investigate in detail how the ACFNC HEA catalyzes the improvement of MgH$_2$ hydrogen storage capabilities. We specifically look into how it affects kinetics and thermodynamics, which sheds light on the mechanisms that underlie its catalytic action. We hope to provide insight into the highly catalytic nature of HEAs in MgH$_2$ systems by our thorough analysis, opening up new possibilities for the advancement of hydrogen storage technology.

2. **Experimental section:**

**2.1. Synthesis of Monophasic BCC Al-Cu-Fe-Ni-Cr High Entropy Alloy (HEA):**

High-purity elemental powders of Al, Cu, Fe, Ni, and Cr were used to start the synthesis. Alfa Aesar supplied the Ni (99.9%), Cerak Speciality Inorganics provided the Cu (99.9%), Riedel-de Haën provided the Fe (99.8%), and Alfa Aesar provided the Al (99.9%) and Cr (99.9%). In order to avoid contamination and to make the alloying of the powders simpler and more effective, hexane was employed as a medium during the mechanical alloying process. In order to synthesize equiatomic Al-Cu-Fe-Ni-Cr (ACFNC) HEA, 20% of each of the primary elements were utilized. A high-energy attritor ball mill running at 400 rpm and with a 40:1 ball-to-powder ratio was used for synthesis of HEA. For mechanical alloying, 7 mm-diameter stainless steel balls were used. The 1.5-liter mixing drum and water jacket in the high energy attritor ball miller help to keep the milling temperature constant. We obtained a single-phase BCC HEA with a lattice parameter of 0.289 nm following the 40-hour milling process. HEA was employed in $MgH_2$ as a catalyst in research on hydrogen de- and re-hydrogenation.

**2.2 Synthesis Pathway for Catalyzed $MgH_2$: Experimental Details and Procedure:**

The starting material used in this study was 99% pure magnesium hydride ($MgH_2$), sourced from Alfa Aesar. A lab-made stainless-steel vial capable of withstanding 100 atm of pressure is utilized for mechanical milling. To fill the vial, 1 gram of $MgH_2$ was mixed with 5 wt. % and 7 wt. % of the produced catalyst ACFNC HEA. After that, the sample was milled for 24 hours at 250 rpm in a hydrogen environment (5 atm) with a 50:1 ball to powder ratio. Various sized stainless-steel balls were employed for the homogenous mixing. To avoid moisture contamination and the production of magnesium oxide, gently replace the hydrogen pressure inside the vial. To keep the levels of $H_2O$ and $O_2$ below 0.5 ppm, all sample handling operations were performed in a glove box (mBRAun MB 10 compact) that operates in a $N_2$ atmosphere. The purpose of this regulated environment was to stop contamination and oxidation. To facilitate a comparative investigation, the catalytic impact of ACFNC HEA on the hydrogenation and dehydrogenation kinetics of $MgH_2$ was thoroughly examined. Notably, this study did not involve any activation studies.

**2.3 Characterization tools:**

We have characterized the synthesized samples using a variety of instruments to gather structural and morphological data. The Pananlytical Empyrean XRD system employed high resolution area detectors measuring 256 x 256 pixels at 40V and 40mA to obtain structural information. In-depth microstructure exploration was conducted using the Technai 20 G$^2$ TEM operating at 200 kV. The surface morphology, topography, and external characteristics were examined using a FEA Quanta 200 SEM operating at 25 kV and 10$^{-6}$ torr vacuum. Energy-Dispersive X-ray Spectroscopy (EDS) in combination with Transmission Electron Microscopy (TEM) and Scanning Electron Microscopy (SEM) was used to evaluate the elemental composition.Chemical analysis was conducted utilizing the Physical Instruments (PHI) 500 Versa Probe-3 spectrometer and Al K alpha radiation (1486.6ev) operator ultra-high vacuum $10^{-12}$ torr. The approach yields detailed information about the chemical composition, electronic structure, and binding properties of materials.Using Pressure-Composition Isotherm (PCI) equipment provided by Advanced Materials Corporation (Pittsburgh, USA), 200 mg of the prepared sample was examined for each hydrogen sorption measurement. Under controlled circumstances, the samplesabilities to absorb and desorb hydrogen were assessed using the PCI system.By enumerating every acronym utilized, Table 1 streamlines this technical investigation and facilitates readers' navigation of the content.

The following formulas were used to determine the thermodynamic parameters that are essential for the synthesis of HEA: mixing entropy ($\Delta S_{mix}$), mixing enthalpy ($\Delta H_{mix}$), atomic size difference ($\delta$), and valence electron concentration (VEC) [33]:

$$\Delta S_{mix} = -R \sum C_i \ln C_i \quad (1)$$

where R is a gas constant and $C_i$ is molarity.

$$\Delta H_{mix} = \sum_{i=1, i \neq j}^{n} \Omega_{ij} C_i C_j \quad (2)$$

$$\Omega_{ij} = 4\Delta^{AB}_{mix}$$

where $\Delta^{AB}_{mix}$ is the mixing enthalpy of binary liquid AB alloys.

$$\delta = 100\sqrt{\sum_{i=1}^{n} C_i(1 - \frac{r_i}{\bar{r}_i})^2} \quad , \bar{r} = \sum_{i=1}^{n} C_i r_i, \quad (3)$$

where $r_i$ and $C_i$ stand for atomic radius and atomic percentage, respectively.

$$VEC = \sum_{i=1}^{n} C_i(VEC)_i, \quad (4)$$

where $(VEC)_i$ stands for the valence electron concentration of the i-th component.

Using these formulas, we synthesized the Al-Cu-Fe-Ni-Cr HEA with the following calculated thermodynamic parameters: $\Delta S_{mix}$ = 13.38 J/mol·K, $\Delta H_{mix}$ = -4 kJ/mol, $\delta$ = 4.9%, VEC = 7.6. These values supported the formation of a single-phase BCC structure in the synthesized alloy.

## 3. Results and Discussion:

### 3.1. An Overview of Structural, Microstructural, and Surface Morphology Traits:

The XRD analysis in Fig. 1 provides fascinating new information about a variety of materials. The HEA XRD pattern, shown in Fig. 1(a), clearly shows a single-phase BCC crystal structure, with a lattice parameter of 2.89Å and a strong peak at 2θ = 44.55°, which corresponds to the (110) plane. The XRD pattern of ball-milled (BM) $MgH_2$ in Fig. 1(b) reveals a well-established tetragonal lattice structure that is part of the space group P42/mnm (136). Its lattice parameters are a = b = 4.516 Å and c = 3.020Å (JCPDS no. 740934). The XRD patterns for the dehydrogenated (Mg + 5 wt. % HEA) and rehydrogenated ($MgH_2$ + 5 wt. % HEA) samples are shown in Figs. 1(c) and 1(d), respectively. These patterns show diffraction peaks coming from $MgH_2$ tetragonal phase as well as its HEA.Significantly, the XRD peaks ofthe HEAinthe dehydrogenated and rehydrogenated samples show a shift towards lower diffraction angles, suggesting changes in the crystal structure brought on by the formation of metal hydrides. Broad diffraction peaks are visible in the HEA milled with $MgH_2$ samples, which are suggestive of nano-sized crystallites and lattice strain brought on by high-energy ball milling. Crucially, as Figs. 1(c) and 1(d) show, HEA phase stability holds steady throughout the re/dehydrogenation process, highlighting its exceptional stability in these circumstances. We carried out EDS analysis and TEM characterization on the $MgH_2$ – 5 wt. % HEA composite to gain a better understanding of the behavior of the catalyst buried inside the matrix. We provide bright field pictures of the $MgH_2$ – 5 wt. % HEA at different resolutions in Figs. 2(a) and 2(c). Figure 2(a) shows a scattering of dark areas. Through EDS analysis, the HEA was determined to be the cause of these dark patches. Additionally, the EDS characterization showed that elemental signals like Al, Cu, Fe, Ni, and Cr overlapped, indicating that the Al-Cu-Fe-Ni-Cr alloy was stable in its distribution within the $Mg/MgH_2$ matrix. The $MgH_2$ – 5 wt. % HEA sample was remarkably free of impurity components, confirming the composite effective production. The SAED ring pattern of $MgH_2$ – 5 wt. % HEA is shown in Fig. 2(b), which sheds light on the material crystalline structure. The presence of nanocrystalline HEA catalyst particles within the composite material is confirmed by the SAED ring pattern, which is indicative of polycrystallinity. An HRTEM image of the $MgH_2$ (111) planes with an inter-planar separation of 0.17 nm is shown in Fig. 2(c). This highlights the material structural integrity by providing proof of its crystalline identity. Our comprehensive TEM and EDS investigations shed light on the microstructural and compositional characteristics of the $MgH_2$-5 wt. % HEA composites, offering crucial details regarding the material phase distribution and crystalline properties.

The structural evolution of the as-synthesized HEA and the MgH$_2$–5 wt. % HEA samples in their rehydrogenated and dehydrogenated states can be seen in detail in the SEM images shown in Figs. 3(a), 3(b), and 3(c). The morphological and microstructural changes that take place during the rehydrogenation and dehydrogenation process are visible in these images. The as-synthesized HEASEM micrograph (Fig. 3(a)) displays a structure like a crumbly crust, which is indicative of mechanical alloying. The changes shown in these micrographs demonstrate how the composition of these complex structures and processing conditions interact dynamically. In Figs. 3(d), 3(e), and 3(f), the elemental composition of the as-synthesized HEA, as well as the rehydrogenated and dehydrogenated samples, is further examined using EDS.This research confirms the equiatomic ratio of principal elements in the as-synthesized HEA and sheds light on the distribution and existence of important components. Crucially, the EDS data highlight the HEA catalyst crucial role in hydrogen storage dynamics by confirming its existence in the MgH$_2$ matrix. The dehydrogenated sample (Mg-5 wt. % HEA) has a noticeable crack in Fig. 3(c), which is explained by the phase change that occurs during dehydrogenation. Fissures emerge in the material as a result of the mechanical stresses caused by this change, in which MgH$_2$ returns to metallic Mg. These results open the door for better-informed design strategies by offering insightful information about how to maximize the performance and dependability of MgH$_2$-HEA composites in hydrogen storage applications.

3.2 **XPS spectroscopy analysis:**

We used XPS to analyses the elemental content and chemical structure of both dehydrogenated and rehydrogenated materials. Comprehensive examinations of the acquired spectra yielded significant insights into the electronic configuration of the catalyst employed in our experimental investigation. Figure 4 and 5 displays the XPS spectra following deconvolution for the rehydrogenated and dehydrogenated materials. The observed changes in the electronic states clearly demonstrate the catalytic involvement of HEA during the hydrogenation and dehydrogenation of $MgH_2$. The $MgH_2$–5 wt. % HEA system primary elements oxidation statuses changed, as shown by the XPS spectrum study. During both rehydrogenation and dehydrogenation, the XPS spectra for Al showed a +3-oxidation state with a single peak at 87 eV in the Al 2P core level [34-35]. This peak did not change, suggesting that the binding energy did not change during the procedures. During rehydrogenation, Cu showed two peaks at 976 eV ($2P_{1/2}$) and 968 eV ($2P_{3/2}$), which represent the +2 and +3 oxidation states, respectively [36-37]. There was only one peak during dehydrogenation, which corresponded to the +2-oxidation state at 976 eV ($2P_{1/2}$). This implies that the reduction of copper oxide to a lower oxidation state, mainly Cu (+2), is a necessary step in the dehydrogenation process [38]. Fe did not change its electronic states during the dehydrogenation or hydrogenation processes. About 710 eV ($2P_{3/2}$) and 721 eV ($2P_{1/2}$), which correspond to the +3 and +2 oxidation states, respectively, were the binding energies for the two peaks of Fe [39-40]. Ni displayed a single peak at roughly 851 eV (2P), which are the +3 oxidation states [41-42]. The process of hydrogenation and dehydrogenation did not cause any changes to this peak. During the +3-oxidation state, Cr showed solitary peaks at 576 eV ($2P_{3/2}$) during rehydrogenation and 574 eV ($2P_{3/2}$) during dehydrogenation [43]. An increase in electron density surrounding the Cr atom is suggested by the binding energy's reported drop following dehydrogenation. Mg displayed a single peak at 87 eV that did not change when the element was rehydrated or dehydrated, suggesting that these procedures had no effect on Mg electronic state [44-45].Overall, the observed changes in the electronic states of several elements in the $MgH_2$–5 wt. % HEA system show how the hydrogenation and dehydrogenation processes are influenced by the HEA catalyst. This emphasizes how important HEA are as catalysts for raising the effectiveness of $MgH_2$-based hydrogen storage systems.

**3.3 De/re-hydrogenation kinetics study of $MgH_2$- LHEA:**

We carried out thorough non-isothermal dehydrogenation tests utilizing $MgH_2$ – 5 wt. % HEA and $MgH_2$ – 7 wt. % HEA in order to thoroughly evaluate the impact of ACFNC HEA on the hydrogen desorption capabilities of $MgH_2$. The outcomes of these studies are shown in Figure 6. Temperature programmed desorption (TPD) analysis was used, in these studies, with a constant heating rate of 5 °C/min. The $MgH_2 \rightarrow Mg + H_2$ process was first triggered by pure $MgH_2$ between 425°C and 455°C, resulting in a final hydrogen desorption capability of 5.7 wt. % Figure 6(a). The starting temperature for hydrogen desorption was dramatically decreased to 180°C by adding merely 5 wt. % of Al-Cu-Fe-Ni-Cr HEA. By raising the HEA content to 7 wt. %, this desorption range was further extended to a lower temperature range of 175°C to 390°C. Decreases in terminal dehydrogenation temperatures and a reduction in total hydrogen storage capacity were the subsequent outcomes of incorporating ACFNC HEA. In particular, the $MgH_2$ – 5 wt. % HEA and $MgH_2$ – 7 wt. % HEA composites reached their final capacities of 6.1 wt.% and 3.1 wt.% after completing dehydrogenation at 400°C and 390°C, respectively. Specifically, $MgH_2$ – 5 wt. % HEA demonstrated an excellent release of more than 6.1 wt.% $H_2$ at 425°C, outperforming the activity of pure $MgH_2$, highlighting the remarkable catalytic ability of ACFNC HEA. TPD analysis further revealed that, in comparison to untreated $MgH_2$ samples, $MgH_2$ – 5 wt. % HEA and $MgH_2$ – 7 wt. % HEA showed lower desorption temperatures of 245°C and 250°C, respectively. This demonstrates the $MgH_2$–ACFNC HEA catalyst great potential for raising the effectiveness of hydrogen storage applications. The absorption and desorption capacities of $MgH_2$ – 5 wt. % HEA and $MgH_2$ – 7 wt. % HEA were assessed in additional studies into rehydrogenation kinetics (shown in Figures 7(a) and 7(b)) at temperatures of 280°C, 300°C, and 320°C with a hydrogen pressure of 15 atm. At 320°C, $MgH_2$ – 5 wt. % HEA achieved a maximum hydrogen absorption of 7.3 wt.% within 3 minutes, whereas $MgH_2$ – 7 wt. % HEA absorbed 6.6 wt.% in just 2 minutes. This demonstrates the superior rehydrogenation kinetics of $MgH_2$ – 5 wt. % HEA under comparable temperature and pressure conditions. As shown in Figures 8(a) and 8(b), the dehydrogenation experiments were carried out at 280°C, 300°C, and 320°C with a hydrogen pressure of 0.1 atm. The findings show that at these temperatures, the $MgH_2$ –5 wt. % HEA composite produced hydrogen at rates of 4 wt. %, 5 wt. %, and 5 wt. %. Interestingly, $MgH_2$ –5 wt. % HEA demonstrated outstanding kinetics, completing almost total hydrogen desorption at 320°C in under 5 minutes. Likewise, at 280°C, 300°C, and 320°C, the $MgH_2$–7 wt. % HEA composite had hydrogen desorption rates of 3.9 wt. %, 4 wt. %, and 4.1 wt. %, respectively. Additionally, it showed quick kinetics, desorbing almost entirely at 320°C in under five minutes. These results demonstrate the effectiveness of both composites, with the $MgH_2$–5 wt. % HEA performing somewhat better, especially at lower temperatures. These results highlight the dual-function catalytic property of ACFNC HEA, which efficiently promotes the hydrogen absorption and desorption events in $MgH_2$. The re/dehydrogenation kinetics study clearly indicates that $MgH_2$ catalyzed by 5 wt. % ACFNC HEA shows significantly faster hydrogen uptake and release than $MgH_2$ –7 wt. % HEA, making it the best option for applications needing quick cycles of hydrogen absorption and desorption.

**3.4 Calculating the Enthalpy Change (ΔH) using Pressure Composition Isotherms (PCI):**

We concentrated on the full thermodynamic characterization of catalyzed $MgH_2$, specifically the enthalpy shift, after performing re/dehydrogenation. Figures 9(a) and 9(b) display the PCI isotherms for $MgH_2$ - 5 wt. % HEA and $MgH_2$ - 7 wt. % HEA, respectively. The matching Van't

Hoff plots, which are utilized to precisely calculate the enthalpy fluctuations, are shown in Figures 9(c) and 9(d) [46-47]. For $MgH_2$ - 5 wt. % HEA, it was found that the enthalpies of absorption and desorption were roughly 75.5 kJ/mol and 78.18 kJ/mol (±5.2), respectively [48]. The values were 35.6 kJ/mol and 48.1 kJ/mol (±5.2) for $MgH_2$ - 7 wt. % HEA. The reciprocal of temperature (1/T) and the natural logarithm of pressure (lnP) are correlated by the Van't Hoff plots shown in Figures 9(c) and 9(d).Interestingly, when comparing virgin $MgH_2$ to $MgH_2$ - 5 wt. % HEA, the enthalpy stays mostly unaltered. The $MgH_2$ - 5 wt. % HEA system is stable, and this means that no new phase formation happened during the catalytic activities. In TPD studies, the desorption temperature decreased dramatically from around 425°C to roughly 180°C. This indicates non-equilibrium conditions caused by catalyst presence and particle size reduction, among other variables. In contrast, the enthalpy of reaction, which is the intrinsic thermodynamic energy under equilibrium conditions, is obtained from PCI measurements. This outcome demonstrates how well the $MgH_2$ - 5 wt. % HEA system works, permitting a significant drop in desorption temperature without sacrificing the system overall enthalpy characteristics. Our thermodynamic analysis emphasizes $MgH_2$ - 5 wt. % functionality and stability.$MgH_2$ - 5 wt. % is a feasible solution for applications needing enhanced kinetics and beneficial thermodynamic features in hydrogen storage systems because of the small change in enthalpy and the large decrease in desorption temperature.

## 3.5 Cyclic stability:

For hydrogen storage materials to be used in practical applications, cyclic stability is essential. Because of its remarkable cyclic stability, $MgH_2$ is a material of note for hydrogen storage. Consequently, a thorough examination of the catalyzed $MgH_2$ samples cyclic stability is necessary for their actual use. $MgH_2$ with 5 wt. % HEA underwent isothermal dehydrogenation (at 0.1 atm hydrogen pressure) and rehydrogenation (at 15 atm hydrogen pressure) experiments, carried out 25 times at 320°C, to confirm the cyclic stability. The hydrogenation (0.05 wt. %) and dehydrogenation (0.03 wt. %) performance are slightly lower in Figure 10. The expansion of the nanoparticle $MgH_2$/Mg is responsible for this decrease in hydrogen storage performance during absorption and desorption procedures. During the cyclic process, it is normal for metal hydrides to aggregate and expand into larger particles, which results in the creation of unreacted magnesium at the center [49]. An incomplete reaction of magnesium in the center might arise from the creation of hydride on the surface, which would impede the diffusion of $H_2$ atoms. The reaction between magnesium and hydrogen during hydrogenation is another reason for the decrease in kinetics and capacity. This happens as a result of $MgH_2$ ($1.5 \times 10^{-16}$ m²/s) much lower H diffusion coefficient than Mg ($4 \times 10^{-14}$ m²/s)[32]. The material ability to maintain stability and functionality over numerous cycles of dehydrogenation and rehydrogenation is shown by the cyclic stability curves.

A comparison of the study findings with earlier research, in which various alloys were used as catalysts to enhance $MgH_2$ for hydrogen storage, highlights the significance of this work. Table 2 provides specifics on these comparisons.

3.6 **Mechanism of catalyst:**

The hydrogen storage performance of $MgH_2$ is significantly improved by the Al-Cu-Fe-Ni-Cr (ACFNC) HEA; however, the underlying catalytic mechanisms need to be thoroughly clarified. The microstructural evolution of $MgH_2$–5wt. % HEA composites was examined using XRD and TEM methods, providing insight into the catalytic mechanisms involved. Critical insights were provided by the XRD patterns (Figs. 1(c) and 1(d)) of the $MgH_2$–5wt. % HEA composites in their dehydrogenated and rehydrogenated phases. $MgH_2$ was clearly visible in the ball-milled condition, showing a clear peak at 44.55°, which corresponds to the ACFNC HEA phase. Crucially, the lack of other peaks suggested that there was no chemical reaction between $MgH_2$ and the ACFNC HEA during the preparatory stage. $MgH_2$ was converted to Mg by dehydrogenation, and then back to $MgH_2$ by cycles of rehydrogenation. The constant peak at 44.55° during these processes indicates that the ACFNC HEA phase stayed stable, indicating that its crystal structure was preserved during hydrogen absorption and desorption. Detailed microstructural and compositional studies were obtained using TEM in conjunction with EDS (Fig. 2(a-d)). $MgH_2$ crystal planes (111), (202), and (222) were validated by SAED patterns (Fig. 2(b)), along with diffraction rings that corresponded to the ACFNC HEA, particularly with a lattice spacing of 0.2 nm. The presence of $MgH_2$ with a lattice spacing of 0.17 nm was further underlined by high-resolution TEM (HRTEM) images (Fig. 2(c)). The composite structure Al, Cu, Fe, Ni, and Cr were found to be uniformly distributed and closely integrated, as shown by the EDS analysis (Fig. 2(d)). The hydrogen diffusion at $Mg/MgH_2$ interfaces was accelerated by the close contact between ACFNC HEA and $MgH_2$ surfaces, which promoted synergistic catalytic actions. Changes in the surface morphology following dehydrogenation and rehydrogenation were demonstrated by SEM-EDS analysis (Fig. 3). The near proximity of Al, Cu, Fe, Ni, and Cr to $MgH_2$ suggested alloying or strong interactions, which are essential for the features of hydrogen storage [56]. By serving as efficient catalytic sites, these components dramatically lower the activation energies required for the processes of hydrogenation and dehydrogenation. XPS analyses of $MgH_2$–5wt. % HEA samples that had been rehydrogenated and dehydrogenated (Figs. 4 and 5) showed that the multivalent environment was favorable to faster re/de-hydrogenation rates. During dehydrogenation, this environment destabilizes Mg-H bonds, making electron transfers easier [57]. ACFNC HEA catalytic function in solid-state hydrogenation reactions is further supported by its ongoing stability throughout cycling processes.The ACFNC HEA complex catalytic mechanism improves the performance of $MgH_2$ hydrogen storage. Effects of Synergistic Alloy: Because of their comparable atomic sizes, the Al, Cu, Fe, Ni, and Cr constituents of the HEA structure form a stable solid solution, which enhances performance [58]. Improved diffusion and reaction rates: several heterogeneous activation sites and diffusion channels are provided by close contact between the ACFNC HEA and $MgH_2$ surface, which is essential for effective hydrogen absorption and desorption [59]. Activation of Hydrogen Bonds: By lowering energy barriers, the inclusion of several catalytic components improves total hydrogen storage kinetics and allows for speedier reaction rates [60].The synergistic effects of Al, Cu, Fe, Ni, and Cr in the ACFNC HEA have been clarified using integrated XRD, TEM-EDS, SEM-EDS, and XPS investigations, indicating its promising role as a catalyst in improving $MgH_2$ hydrogen storage technologies. These results highlight the alloy potential to significantly improve the kinetics of hydrogen absorption and desorption, which is important for upcoming hydrogen-based energy applications.

4. **Conclusion:**

This work presents a new class of single-phase BCC ACFNC HEA catalysts that show great promise for improving $MgH_2$ hydrogen storage characteristics, especially when it comes to overcoming kinetics and thermodynamics-related issues. Merely adding 5 wt. % of HEA to $MgH_2$ greatly improves its re/dehydrogenation properties. Interestingly, $MgH_2$– 5 wt. % HEA significantly lowers its dehydrogenation temperature to about 180°C, a 245°C drop from pristine $MgH_2$. $MgH_2$–5 wt. % HEA shows efficient desorption (~5 wt. % in 6 minutes) and quick hydrogen absorption (~7.3 wt. % in approximately 5 minutes) at 320°C and 15 atm hydrogen pressure. Furthermore, over a span of 25 cycles, the HEA catalyst exhibits strong cyclic stability, with minor losses in hydrogen storage capacity (~0.05 wt. % for re-hydrogenation and ~0.03 wt. % for dehydrogenation). The HEA is essential in reducing the activation energy barriers for hydrogen desorption, even while the thermodynamic parameter ΔH of $MgH_2$ remains unchanged. This work clarifies the complex behavior of HEA and establishes its critical function as a potent $MgH_2$ catalyst. The encouraging results of this study provide fresh opportunities to apply HEA for catalyzing different hydrides, potentially leading to improvements in hydrogen storage technology.

**Table 1:** Abbreviated names of the processed Al-Cu-Fe-Ni-Cr sample.

| S. No. | Sample name | Abbreviated name |
|---|---|---|
| 1 | Al-Cu-Fe-Ni-Cr high entropy alloy | ACFNC HEA |
| 2 | Pristine ball-milled $MgH_2$ | BM $MgH_2$ |
| 3 | $MgH_2$ catalyzed by 5wt. % Al-Cu-Fe-Ni-Cr high entropy alloy | $MgH_2$- 5wt. % HEA |
| 4 | $MgH_2$ catalyzed by 7wt. % Al-Cu-Fe-Ni-Cr high entropy alloy | $MgH_2$–7wt. % HEA |

**Table 2:** A comparison of the hydrogen storage behaviors of $MgH_2$ catalyzed by different transition metal alloys.

| S. No. | Material composition | Processing Method | Onset dehydrogenation temperature(ºC) | Hydrogen storage Capacity (wt.%) | Hydrogen release kinetics | Cyclic stability | Ref. |
|---|---|---|---|---|---|---|---|
| 1 | $MgH_2$ + 5% FeCoNiCrMn | As cast | 275 | 6.1 | 4.5 wt. % in 40 min | 50 cycles | [29] |
| 2 | $MgH_2$ + 9% FeCoNiCrMo | As cast | 200 | 6.7 | 5.2 wt. % in 17 min | 20 cycles | [50] |
| 3 | MgH2 + 10% CrFeCoNi | As cast | 232 | 6.5 | 5.6 wt. % in 10 min | 20 cycles | [51] |
| 4 | $MgH_2$ + 50% NiMnAlCoFe | As cast | 180 | 2 | - |  | [52] |
| 5 | $MgH_2$ + 10% TiVNbZrFe | As cast | 200 | 5.78 | 5.25 wt. % in 30 min | 100 | [53] |
| 6 | $MgH_2$ +10% TiCuNiZr | As cast | 250 | 5.5 | 4.9 wt. % in 5 min | - | [54] |
| 8 | $MgH_2$ + 7% LHEA ($Al_{20}Cr_{16}Mn_{16}Fe_{16}Co_{16}Ni_{16}$) | Leached version of HEA | 338 | 6.8 | 5.4 wt. % in 40 min | 25 | [30] |
| 9 | $MgH_2$ + 7% LHEA (AlCuFeNiTi) | Leached version of HEA | 200 | 6.2 | 5.8 wt. % in 3.8 min | 21 | [55] |
| 10 | $MgH_2$ + 5% AlCuFeNiCr | MA | 180 | 7.3 | 5 wt. % in 6 min | 25 | Present study |

59. Julcour, C., Le Lann, J. M., Wilhelm, A. M., & Delmas, H. (1999). Dynamics of internal diffusion during the hydrogenation of 1, 5, 9-cyclododecatriene on Pd/Al2O3. *Catalysis today*, *48*(1-4), 147-159.

60. Luo, Q., Li, J., Li, B., Liu, B., Shao, H., & Li, Q. (2019). Kinetics in Mg-based hydrogen storage materials: Enhancement and mechanism. *Journal of Magnesium and Alloys*, *7*(1), 58-71.

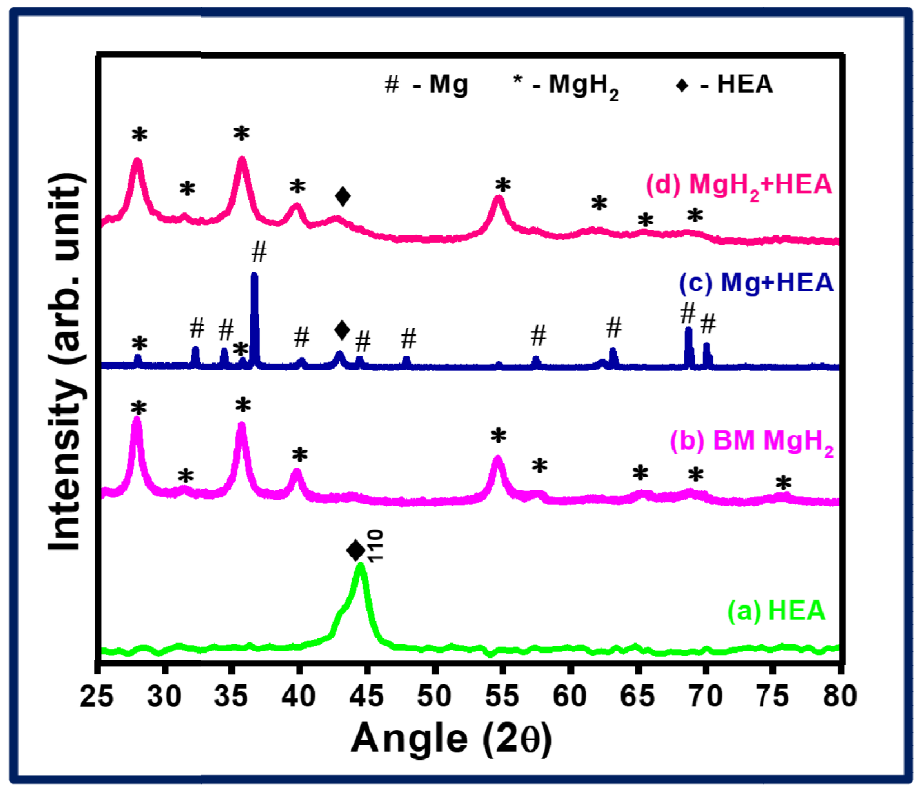

**Fig. 1:** X-ray diffraction patterns depicting (a) as synthesized Al-Cu-Fe-Ni-Cr High Entropy Alloy (ACFNC), (b) ball-milled (BM) MgH$_2$, (c) Mg + 5 wt. % HEA (dehydrogenated sample), and (d) MgH$_2$ + 5 wt. % HEA (rehydrogenated sample).

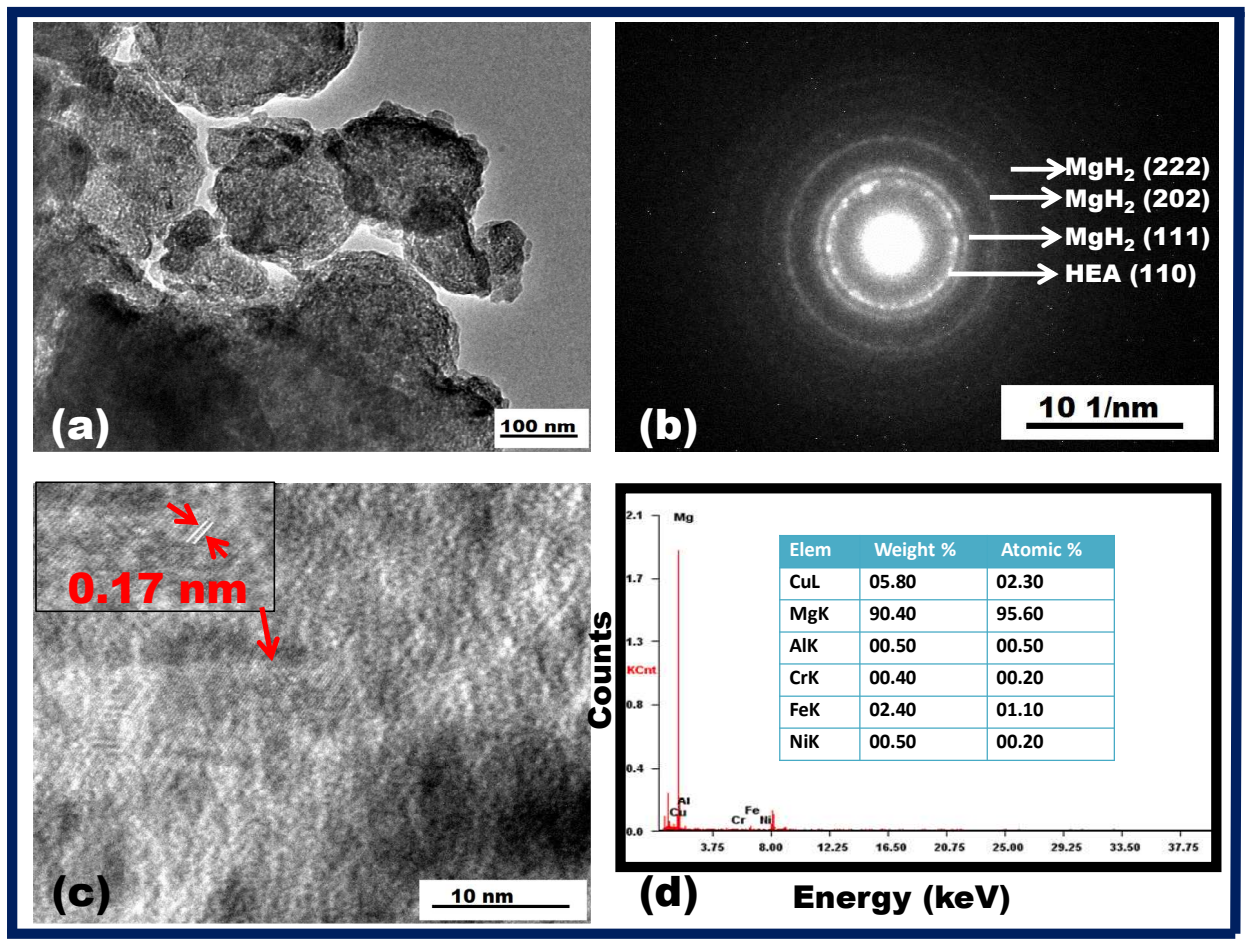

**Fig. 2:** (a) TEM micrograph showing bright field imaging of $MgH_2$-5 wt.% HEA, (b) corresponding SAED pattern, (c) HRTEM image, and (d) EDS spectra with elemental composition analysis.

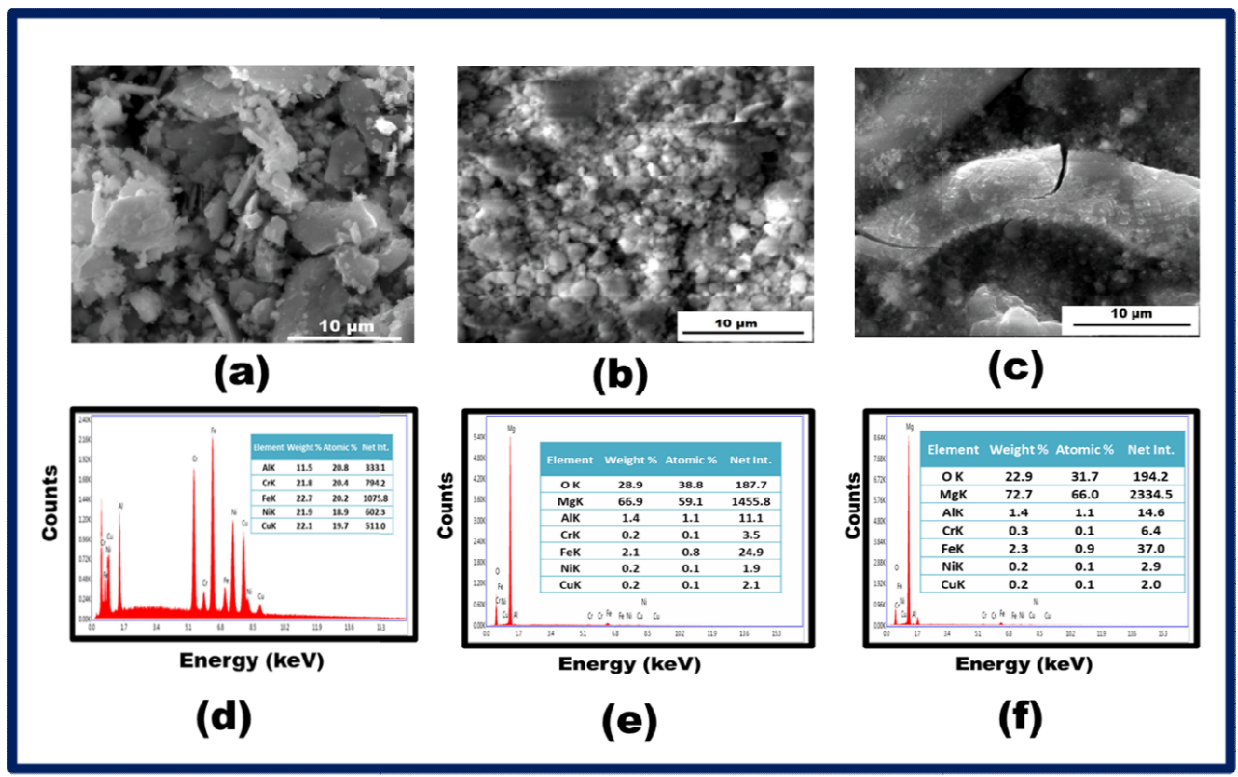

**Fig. 3:** (a) SEM micrograph of the synthesized HEA, (b) SEM micrograph of MgH$_2$ + 5 wt. % HEA (rehydrogenated sample), (c) SEM micrograph of Mg + 5 wt. % HEA (dehydrogenated sample), (d) EDS spectrum showing the elemental composition of the synthesized HEA, (e) EDS spectrum showing the elemental composition of MgH$_2$ + 5 wt. % HEA (rehydrogenated sample), and (f) EDS spectrum showing the elemental composition of Mg + 5 wt. % HEA (dehydrogenated sample).

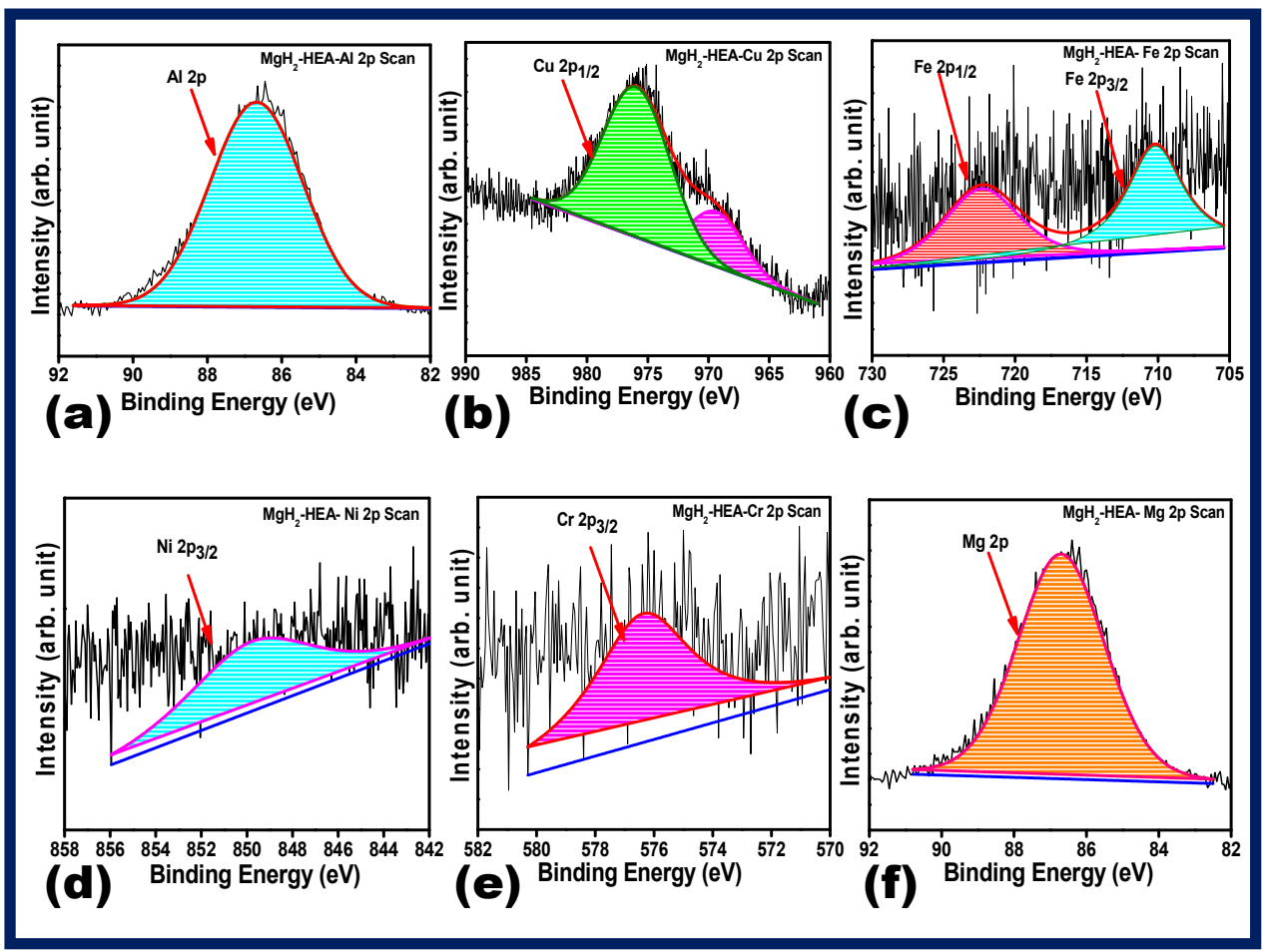

**Fig. 4:** X-ray photoelectron spectroscopy (XPS) spectra showing the 2P regions of catalyst elements in $MgH_2$ + 5 wt. % HEA (rehydrogenated sample).

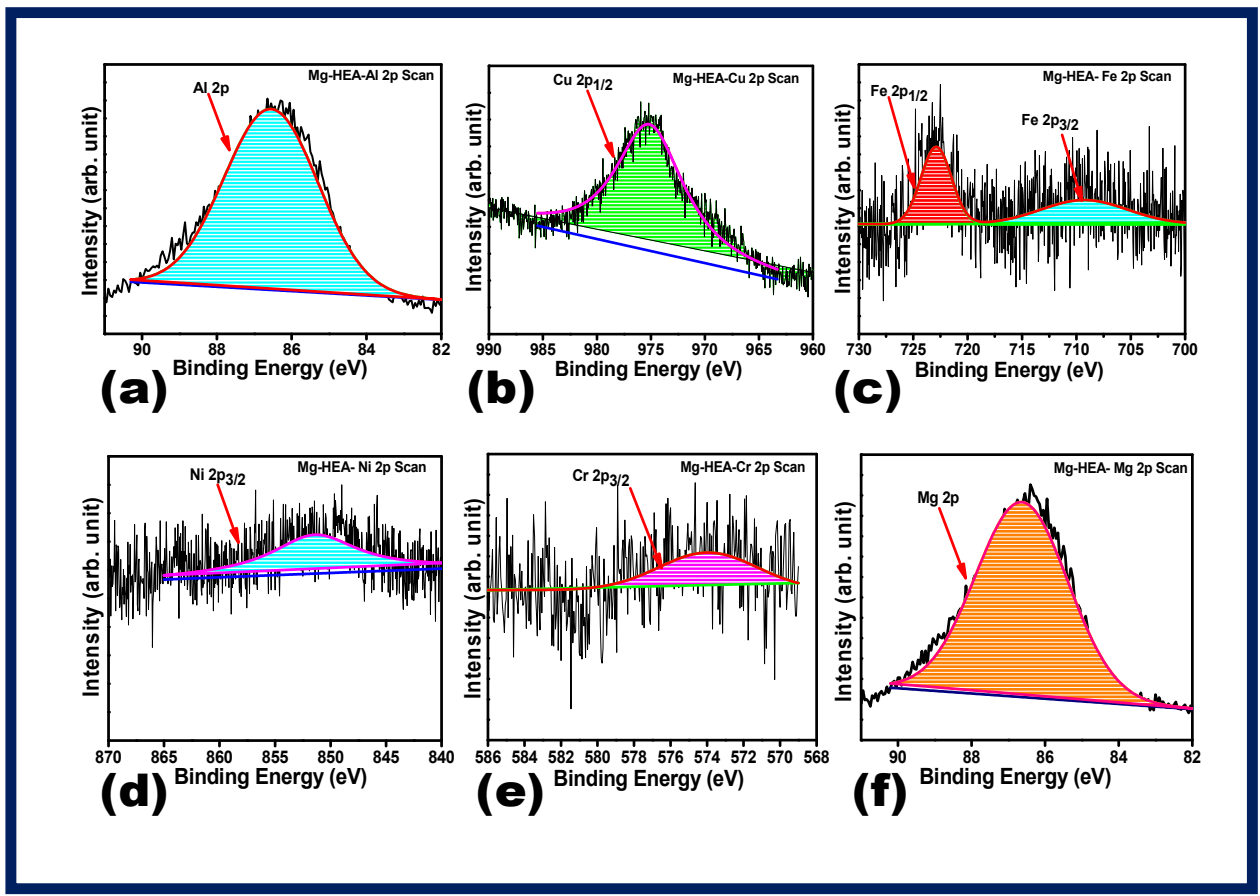

**Fig. 5:** X-ray photoelectron spectroscopy (XPS) spectra showing the 2P regions of catalyst elements in Mg + 5 wt. % HEA (dehydrogenated sample).

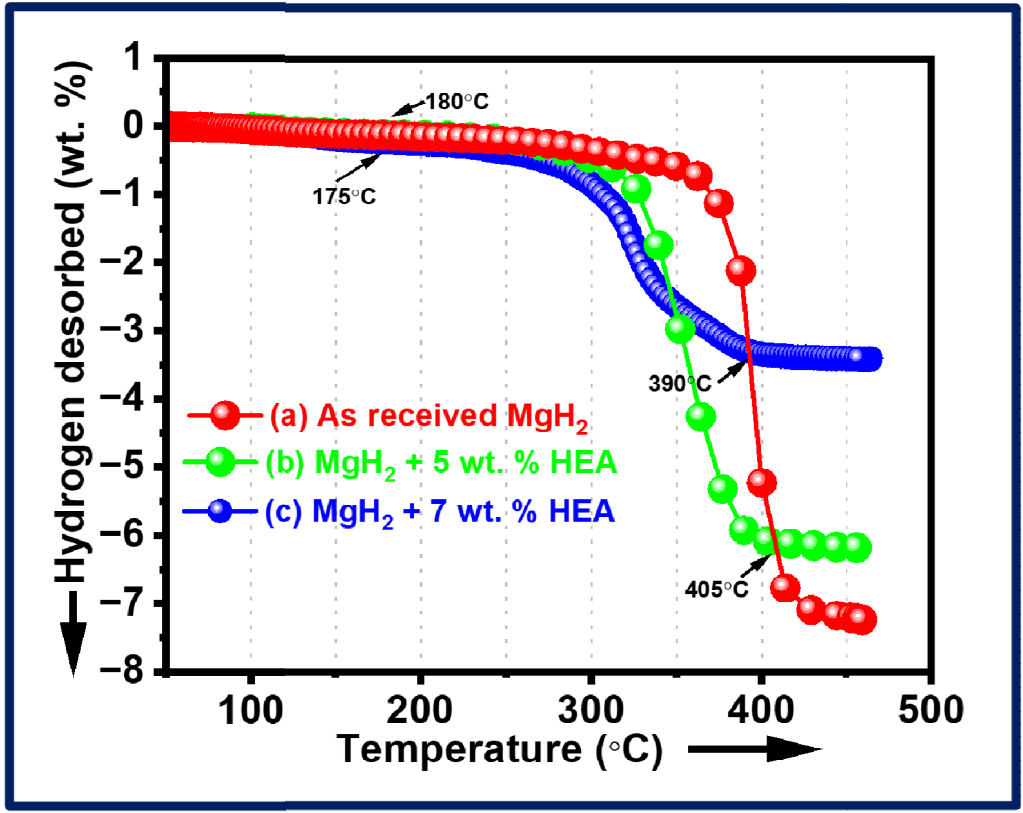

**Fig. 6:** Temperature Programmed Desorption (TPD) curves for (a) Untreated $MgH_2$, (b) $MgH_2$ – 5 wt. % HEA, and (c) $MgH_2$ - 7 wt. % HEA.

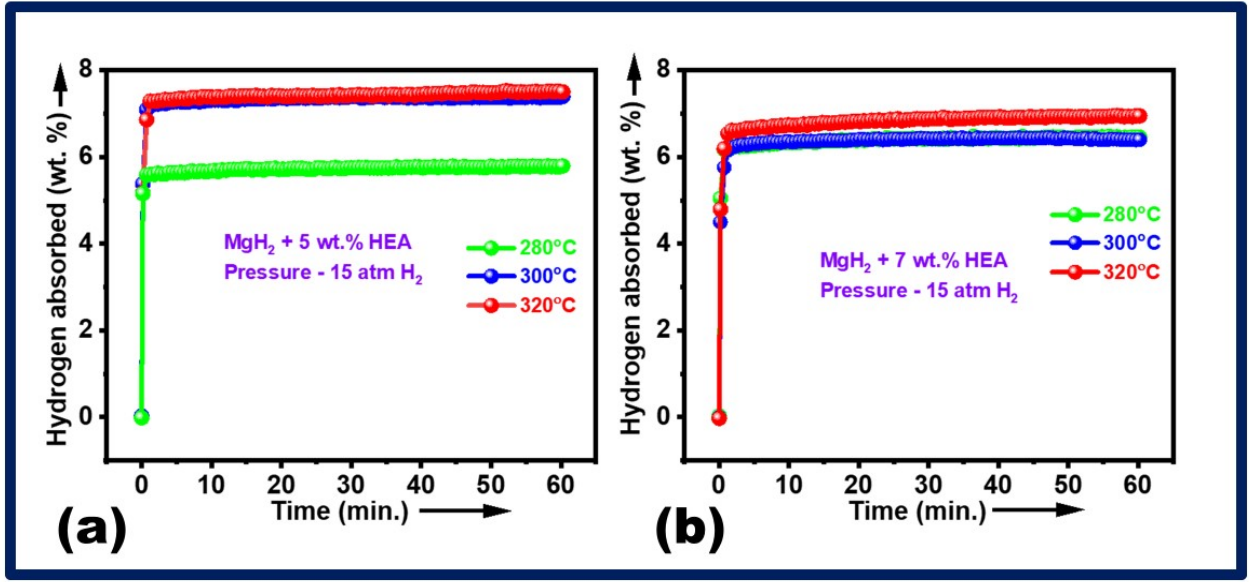

**Fig. 7:** Rehydrogenation curves for (a) MgH$_2$ – 5 wt. % HEA and (b) MgH$_2$ - 7 wt. % HEA at temperatures of 280°C, 300°C, and 320°C under 15 atm H$_2$ pressure.

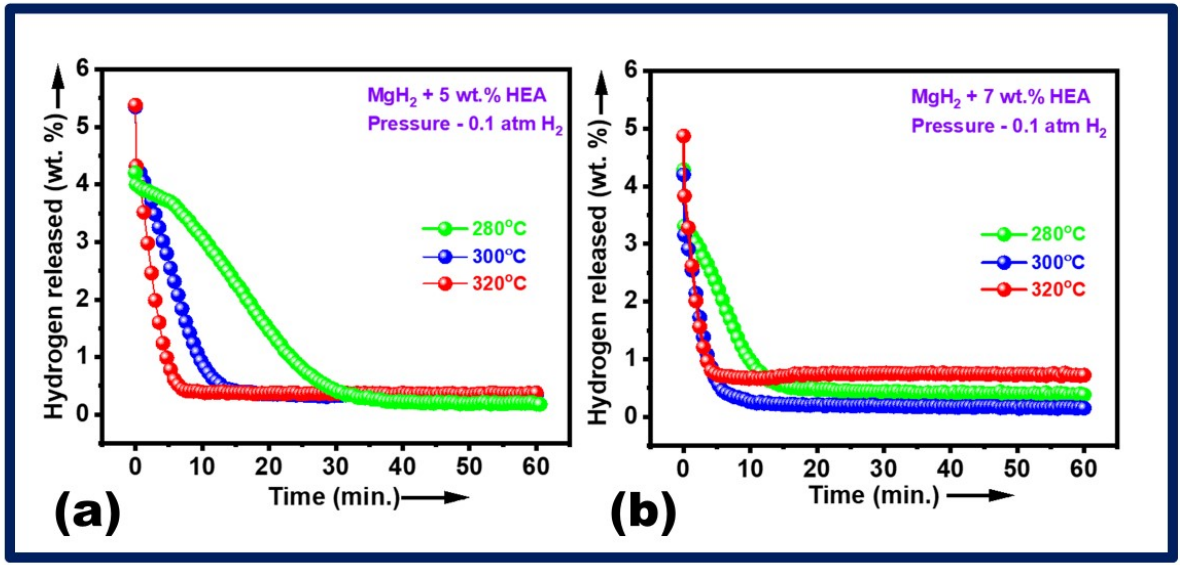

**Fig. 8:** Dehydrogenation curves for (a) $MgH_2$ – 5 wt. % HEA and (b) $MgH_2$ - 7 wt. % HEA at temperatures of 280°C, 300°C, and 320°C under 0.1 atm $H_2$ pressure.

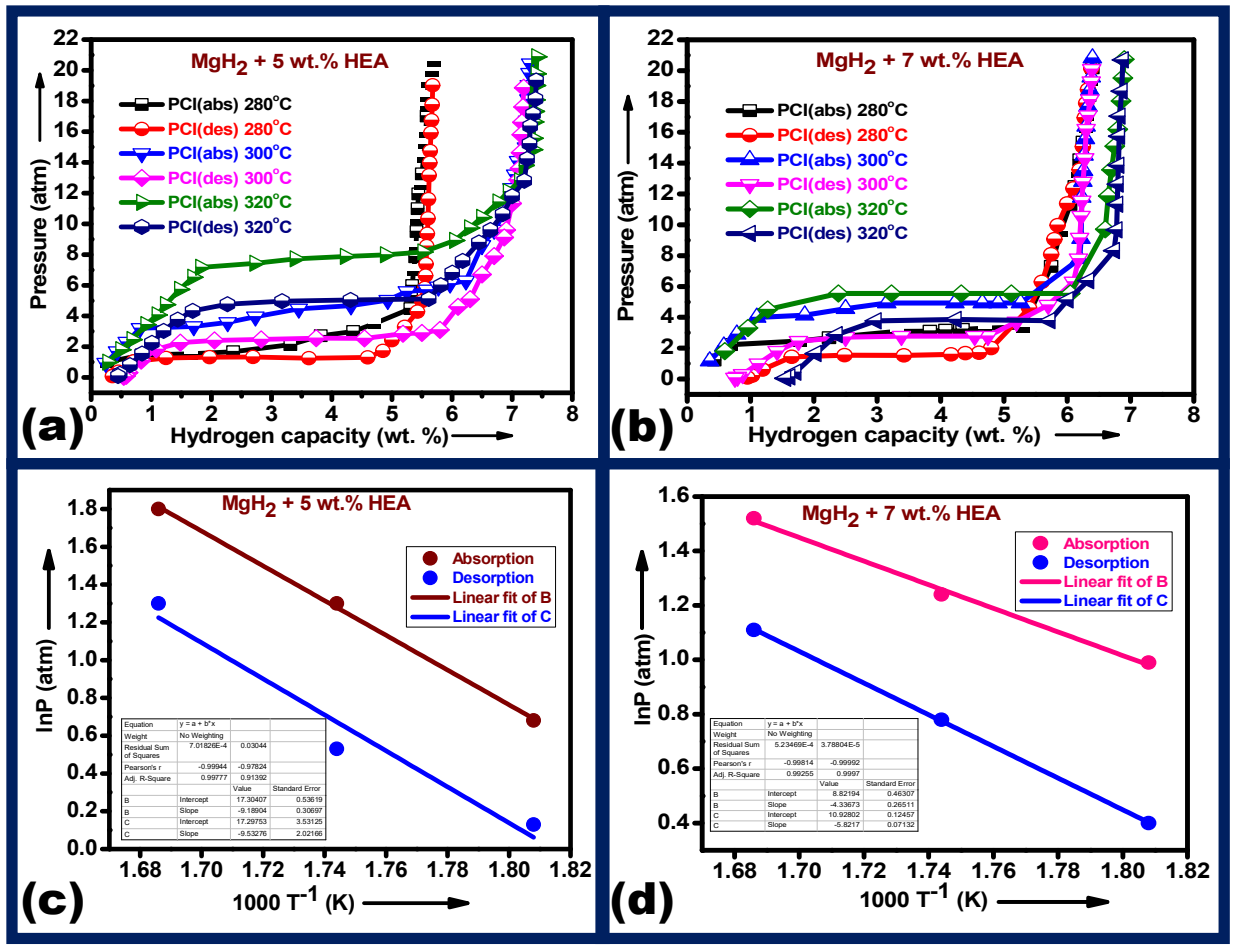

**Fig. 9:** (a) Pressure-composition isotherm (PCI) showing the absorption and desorption of MgH$_2$ with 5 wt. % HEA at three different temperatures. (b) Pressure-composition isotherm (PCI) illustrating the absorption and desorption of MgH$_2$ with 7 wt. % HEA at three different temperatures. (c) Corresponding Van't Hoff plot for MgH$_2$ with 5 wt. % HEA. (d) Corresponding Van't Hoff plot for MgH$_2$ with 7 wt. % HEA.

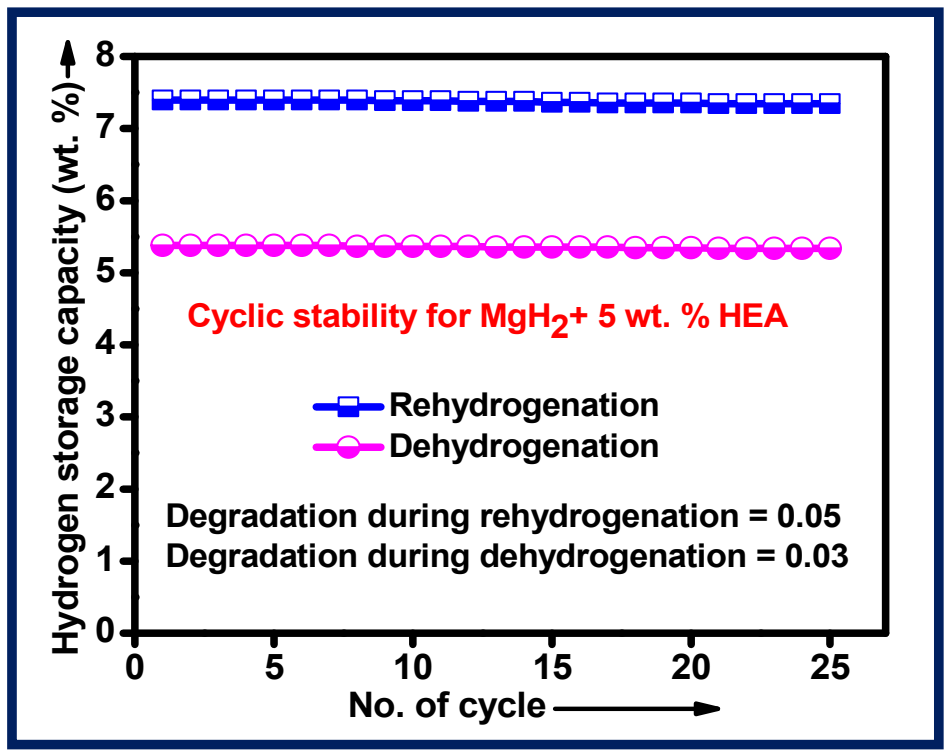

**Fig. 10:** Cyclic stability curves illustrating the rehydrogenation and dehydrogenation behavior of $MgH_2$-5 wt.% HEA over 25 cycles. Rehydrogenation was conducted at 320°C under 15 atm $H_2$ pressure, while dehydrogenation was performed at 320°C under 0.1 atm $H_2$ pressure.